# Machine Learning Pipeline for Software Engineering: A Systematic Literature Review

Samah Kansab

Supervised by Mohammed Sayagh
Co-supervised by Francis Bordeleau

Software and IT Engineering Department,
École de technologie supérieure (ÉTS), Montreal, Canada
samah.kansab.1@etsmtl.net, mohammed.sayagh@etsmtl.net,
francis.bordeleau@etsmtl.net

**Abstract**

The rapid advancement of software development practices has introduced significant challenges in ensuring quality and efficiency across the software engineering (SE) lifecycle. As SE systems grow in complexity, traditional approaches often fail to scale, resulting in prolonged debugging times, inefficient defect detection, and resource-intensive development cycles. Machine Learning (ML) has emerged as a transformative solution, enabling automation and optimization in critical SE tasks such as defect prediction, code review, and release quality estimation. However, the effectiveness of ML in SE relies heavily on the robustness of its underlying pipeline, which includes data collection, preprocessing, feature engineering, algorithm selection, model validation, and evaluation. This systematic literature review (SLR) investigates the state-of-the-art ML pipelines tailored for SE, consolidating advancements and identifying best practices, challenges, and gaps. Our findings reveal that robust preprocessing techniques, such as SMOTE for data balancing and SZZ-based algorithms for feature selection, significantly enhance model reliability. Ensemble methods like Random Forest and Gradient Boosting outperform in most SE tasks, while simpler models, such as Naïve Bayes, offer efficiency and interpretability in specific contexts. Evaluation metrics like AUC, F1-score, and precision are widely adopted for assessing model performance, with emerging metrics such as Best Arithmetic Mean (BAM) showing promise in specialized applications. Validation techniques like bootstrapping ensure model stability and generalizability. This SLR underscores the relevance of ML pipelines in addressing contemporary SE challenges and provides actionable insights for researchers and practitioners aiming to optimize software quality and efficiency. By identifying gaps and emerging trends, this study establishes a foundation for advancing ML adoption in SE and fostering innovation in increasingly complex development environments.

# 1 Introduction

The rapid advancement of software development practices has led to increasingly complex and large-scale software systems. As projects grow in size and complexity, software teams face significant challenges in ensuring quality and efficiency across various stages of the software development lifecycle. For instance, prolonged execution and validation times for tasks such as defect detection, code integration, and testing are common, often leading to inefficiencies and delays in project completion [68, 47, 87]. Debugging pipelines and identifying root causes for failure can take hours, if not days, of developer effort, only to result in pipelines terminating with errors [67]. This raises critical concerns about resource utilization and the scalability of traditional Software Engineering (SE) practices.

To address these challenges, Machine Learning (ML) has emerged as a transformative technology, enabling automation and optimization across various SE tasks. ML has been successfully applied to streamline defect prediction [27, 19], code review [8, 56], software release quality estimation [71], and cost optimization [104]. For example, a study by Li, Niu, and Jing [60] discusses the application of ML methods in defect prediction, supporting developers in handling defects before they are introduced into the production environment. A key enabler of ML's success in SE lies in its ability to process and learn from diverse software artifacts, including source code, test cases, and bug reports [81, 103]. However, leveraging ML for SE is not without its challenges [7, 38]. The effectiveness of an ML model in SE often hinges on the robustness of its underlying pipeline, which consists of critical stages such as data collection, preprocessing, feature engineering, model training, validation, and evaluation. In this paper, we present a systematic literature review to analyze the ML pipeline within the SE domain. Specifically, we investigate how the existing literature addresses each stage of the pipeline, identifying common practices, challenges, and opportunities for improvement.

One prominent area of focus in ML pipelines is data quality. Data preparation is a critical component of machine learning projects, often consuming a significant portion of the total effort. A study by Sancricca, Siracusa, and Cappiello [88] found that data scientists spend approximately 80% of their time on data preparation tasks, underscoring the importance of robust data preprocessing techniques. For instance, handling imbalanced datasets through techniques like SMOTE (Synthetic Minority Oversampling Technique) or advanced resampling methods has been critical in improving model performance for defect prediction [98, 17, 107]. Similarly, feature engineering techniques, such as using SZZ-based algorithms to identify bug-inducing changes in code, have been pivotal in ensuring meaningful feature selection [92].

Selecting the appropriate learning algorithm is another critical challenge in optimizing ML pipelines for SE. Ensemble methods, including Random Forest, Gradient Boosting, ect, have consistently delivered superior results in SE tasks, particularly in defect prediction and code quality estimation [79, 36]. On the other hand, deep learning models such as convolutional neural networks (CNNs) and recurrent neural networks (RNNs) have demonstrated success in processing textual and sequential data, such as commit messages and change logs [65].

Model evaluation and validation introduce additional complexities to the optimization of ML pipelines in SE. While standard metrics such as F1-score, precision, and recall are prevalent and provide valuable insights, it is essential to critically examine their suitability and the impact of relying solely on these metrics across diverse SE tasks. Furthermore, advanced validation techniques, including bootstrapping and cross-validation, have been widely adopted, offering robust mechanisms to assess model performance and generalizability. However, a comprehensive analysis is needed to understand the trade-offs and best practices associated with these approaches, particularly in the context of SE-specific challenges and datasets.

Despite the growing body of research, a unified understanding of ML pipeline optimization for SE remains elusive. Current literature often focuses on isolated aspects of the pipeline, such as data preprocessing or algorithm selection, without providing an end-to-end perspective. To address this gap, this paper conducts this SLR to consolidate and analyze advancements in the ML pipeline for SE. Our review aims to identify best practices, emerging trends, and key challenges across the pipeline stages. Specifically, we address the following research questions:

1. **RQ1: What are the most effective data preprocessing techniques for improving the reliability of ML models in SE?**
   *Key Findings:* Effective preprocessing, including data collection strategies, feature selection, and noise handling, significantly enhances model reliability. Techniques like SMOTE for data balancing and advanced SZZ variants for bug data collection stand out for their impact on defect prediction tasks.

2. **RQ2: Which ML algorithms demonstrate the best performance for SE tasks?**
   *Key Findings:* Ensemble methods, such as Random Forest and Decision Trees, consistently deliver strong results across SE applications, including defect prediction. Deep learning models exhibit superior performance on large datasets, while simpler models like Naïve Bayes remain popular for their interpretability and efficiency in specific tasks.

3. **RQ3: What evaluation metrics most accurately assess the effectiveness of ML models in SE applications, and how do these metrics impact model selection?**
   *Key Findings:* Metrics like AUC, F1, precision, and recall are widely adopted due to their balance between sensitivity and specificity. Emerging metrics, such as Best Arithmetic Mean (BAM), show promise for specialized tasks like feature selection in imbalanced datasets.

4. **RQ4: How do different model validation techniques affect the stability and generalizability of ML models in SE?**
   *Key Findings:* Bootstrapping provides stable and generalizable results across various metrics, while walk-forward validation is particularly effective in scenarios requiring temporal data integrity, such as real-world software release cycles.

By addressing these research questions, this study aims to provide actionable insights for practitioners and researchers seeking to refine ML pipelines for SE. The remainder of this paper is organized as follows: Section 2 details the research methodology, Section 3 presents the results of our SLR, and Section 4 concludes with a discussion of findings and implications for future research.

## 2 Background

### 2.1 Definition of Machine Learning (ML)

Machine Learning (ML) refers to a collection of techniques and algorithms that enable computer systems to learn and adapt to solve problems without being explicitly programmed. Rather than relying on hardcoded instructions, ML systems use data to identify patterns, extract insights, and make predictions or decisions. As a subset of artificial intelligence (AI), ML emphasizes developing models that improve their performance over time through iterative learning processes [51, 54]. Typical ML tasks include:

- **Classification:** Assigning data points to predefined categories.
- **Regression:** Predicting continuous values based on input features.
- **Clustering:** Grouping similar data points into clusters without predefined labels.
- **Dimensionality Reduction:** Reducing the number of variables under consideration while retaining essential information.

### 2.2 ML Pipeline

An ML pipeline refers to the structured sequence of processes involved in developing, deploying, and maintaining an ML model. Each stage in the pipeline plays a critical role in ensuring the accuracy, robustness, and efficiency of the resulting model. The key stages in an ML pipeline are:

- **Data Collection:** This is the foundational step where raw data is gathered from various sources, such as logs, databases, or external APIs. High-quality data collection ensures that the ML model is trained on relevant and representative datasets [86].
- **Data Preprocessing and Feature Engineering:** This combined step ensures the data is suitable for analysis and optimally structured for learning. It involves cleaning and transforming data by handling missing values, removing noise, and standardizing formats through techniques such as normalization, scaling, and encoding. Additionally, feature engineering focuses on selecting or creating relevant features that enhance the model's ability to identify meaningful patterns. This includes feature selection, extraction, and dimensionality reduction to reduce complexity while retaining essential information [72].
- **Model Training:** Using the prepared data, algorithms are trained to optimize specific performance metrics. Various learning techniques, such as supervised, unsupervised, or reinforcement learning, may be employed depending on the problem [89].

- **Model Evaluation:** After validation, models are evaluated using metrics like precision, recall, F1-score, and AUC to assess their effectiveness in solving the task at hand [74].
- **Model Validation:** Validation ensures that the trained model generalizes well to unseen data. Techniques such as k-fold cross-validation, bootstrapping, and holdout methods are used to evaluate model performance and avoid overfitting [106].
- **Model Deployment and Monitoring:** The final step involves integrating the trained model into a production environment. Continuous monitoring is crucial to ensure the model performs well over time, particularly as input data distributions may change.

## 2.3 Software Engineering (SE)

Software Engineering (SE) is the systematic application of engineering principles to the design, development, testing, deployment, and maintenance of software systems. SE focuses on creating reliable, scalable, and maintainable software solutions that meet user requirements and perform efficiently under various conditions. Core activities in SE include requirements analysis, coding, testing, and debugging.

## 2.4 Machine Learning for Software Engineering (ML4SE)

Machine Learning (ML) has emerged as a transformative tool in Software Engineering (SE), addressing a wide range of challenges, including increasing complexity of software systems, scalability requirements, and the demand for improved efficiency. Using data-driven approaches, ML enables the automation and optimization of various SE tasks. Examples of key applications of ML in SE include:

- **Defect Prediction:** Utilizing historical data to identify code regions that are likely to contain defects, enabling proactive measures to improve software quality.
- **Code Review:** Automating the detection of risky changes and providing recommendations to streamline the code review process, thereby enhancing developer productivity.
- **Release Quality Estimation:** Predicting the likelihood of successful software releases by analyzing factors such as test results, code changes, and historical trends.
- **Cost Optimization:** Improving resource allocation and project planning to reduce costs while maintaining or enhancing software quality.

Beyond these applications, ML has been employed in numerous other areas of SE, demonstrating its versatility and potential to revolutionize traditional workflows. By integrating ML techniques into SE practices, organizations can address longstanding inefficiencies and unlock new possibilities for innovation.

## 2.5 ML Pipeline for SE

The ML pipeline for SE adapts the generic ML pipeline to address SE-specific challenges and datasets. This specialized pipeline involves:

- **Data Collection:** Gathering SE artifacts like source code, bug reports, and commit logs.
- **Preprocessing:** Handling imbalanced datasets, cleaning noisy data, and addressing missing values in SE datasets.
- **Feature Engineering:** Extracting meaningful features from software metrics or using domain-specific techniques like SZZ algorithms for bug-inducing changes.
- **Model Training:** Leveraging algorithms suited for SE tasks, such as Random Forest, Gradient Boosting, or neural networks.
- **Validation and Evaluation:** Using metrics like precision, recall, F1-score, and AUC to assess model performance, focusing on SE-specific objectives like defect prediction accuracy.
- **Deployment:** Integrating ML models into SE workflows, such as continuous integration pipelines, with ongoing monitoring to ensure sustained performance.

By tailoring the ML pipeline to SE contexts, researchers and practitioners aim to optimize software development processes, improve quality, and enhance efficiency, paving the way for innovative solutions in the domain.

## 3 Research Method

This section outlines our systematic approach to collecting, selecting, and reviewing the papers that form the foundation of our SLR on ML pipeline for SE. Figure 1 summarizes the paper selection process, detailed in the following subsections.

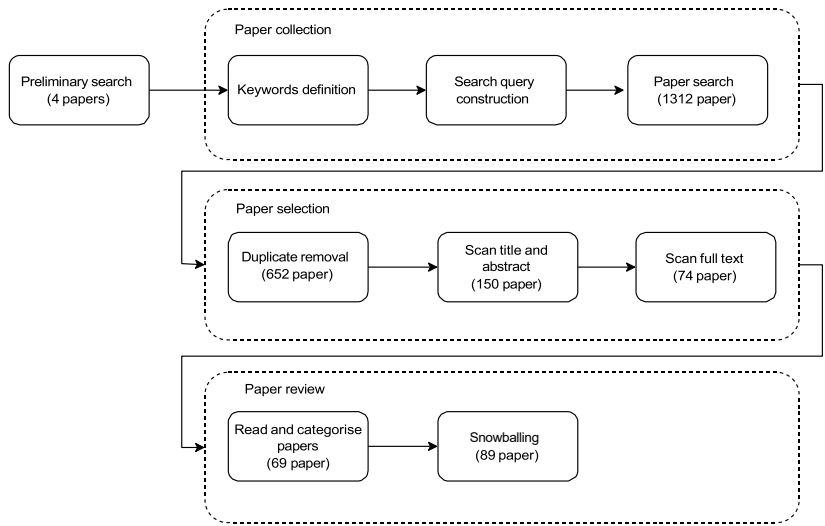

**Fig. 1** Outline of our methodology

## 3.1 Research Questions (RQs)

In this paper, we examine the ML pipeline as shown in Figure 2. Based on this, we design the following RQs to address critical challenges in applying ML to software engineering (SE):

1. **RQ1: What are the most effective data preprocessing techniques for improving the reliability of ML models in SE?**
   *Motivation:* Data preprocessing is often the foundation of a successful ML pipeline. In the SE domain, raw data is frequently noisy, imbalanced, and high-dimensional. Identifying the most effective preprocessing techniques can significantly enhance model performance, reliability, and generalizability, addressing critical issues like imbalanced defect data or noisy metrics.
2. **RQ2: Which ML algorithms demonstrate the best performance for SE tasks?**
   *Motivation:* The diversity of SE tasks, ranging from defect prediction to effort estimation, requires tailored ML solutions. By identifying algorithms that consistently perform well across various SE applications, we aim to guide practitioners in selecting appropriate methods, reducing experimentation overhead, and improving adoption in real-world settings.
3. **RQ3: What evaluation metrics most accurately assess the effectiveness of ML models in SE applications, and how do these metrics impact model selection?**
   *Motivation:* Evaluation metrics are crucial for determining the success of ML models. In SE, traditional metrics like accuracy might not capture the nuances of tasks such as defect prediction or effort estimation. This RQ seeks to identify metrics that align closely with SE goals, ensuring that model selection aligns with both technical performance and practical utility.
4. **RQ4: How do different model validation techniques affect the stability and generalizability of ML models in SE?**
   *Motivation:* Reliable validation is essential for ensuring ML models generalize to unseen data. In SE, where datasets often have limited size or domain-specific characteristics, selecting the right validation strategy is critical. This RQ explores the impact of techniques like cross-validation, hold-out validation, and project-specific validation on model robustness and real-world applicability.

## 3.2 Paper Collection

Given the rapid expansion of ML research in SE, there is a corresponding increase in studies focused on enhancing the ML pipeline for SE. To systematically identify and collect relevant literature, we examine journals and conferences. Our objective is to pinpoint papers that specifically address advancements in the ML pipeline within the SE domain. To ensure a comprehensive and unbiased exploration, we use a structured search strategy using the following databases:

- Engineering Village (Compendex + Inspec) [1]

---

[1]https://www.engineeringvillage.com/home.url

- ACM Digital library [2]
- IEEExplore [3]
- Springer [4]
- Science Direct [5]

### 3.2.1 Keyword Definition

To refine our search and ensure the relevance of retrieved articles, we designed a precise set of keywords through an iterative process:

- **Initial Broad Search:** We begin with general terms which are **"software engineering"**, **"machine learning"**, and **"prediction OR classification model"**, aiming to capture a wide array of studies. To ensure that the retrieved results are specifically pertinent to SE, we carefully filter the sources to include only those explicitly related to SE domains, thereby maintaining the relevance and precision of our review.
- **Contextual Focus:** By reviewing the following highly cited papers on ML pipeline for SE Tantithamthavorn, Hassan, and Matsumoto [98], Jiarpakdee, Tantithamthavorn, and Treude [45], Jiarpakdee, Tantithamthavorn, and Hassan [44], and Lin, Tantithamthavorn, and Hassan [62], we observe recurring terms such as **"analytical"** and **"interpret*"**. These terms are associated with studies focusing on ML's application and impact within SE, which guide us in narrowing our focus.
- **Prominent Researchers:** We analyze publications by leading researchers in SE defect prediction, identifying **"defect"** as a critical keyword. This refinement allow us to target studies on ML applications addressing software defects.
- **Optimization of Search Logic:** To further enhance precision, we exclude terms unrelated to SE and ML, such as **"cloud"**, **"medical"**, and **"image"**. This exclusion help eliminate irrelevant domains from the results.

Based on these refinements, the final query is structured to prioritize studies relevant to ML pipeline for SE, focusing on predictive and interpretive aspects:

**Query:**
(("SE" WN ST) AND ("ML OR model OR defect OR (software AND (model OR ML OR defect OR AI)" WN TI)) AND ("ML OR model OR classifier" AND "(prediction OR classification OR analytical OR interpret*)" WN AB)) NOT (cloud OR medical OR image WN all)

### 3.2.2 Paper Selection

To ensure the relevance and quality of the selected studies, we applied a rigorous inclusion and exclusion process. This filtering aimed to focus on research that directly addresses advancements in Machine Learning (ML) pipelines within the Software Engineering (SE) domain. The inclusion and exclusion criteria are outlined in Table 1.

---

[2]https://dl.acm.org/
[3]https://ieeexplore.ieee.org/Xplore/home.jsp
[4]https://link.springer.com/
[5]https://www.sciencedirect.com/

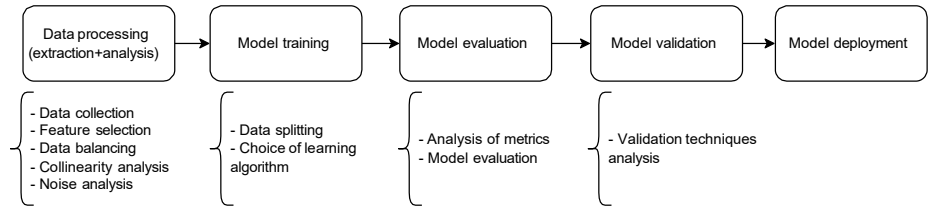

**Fig. 2** ML Pipeline for SE

The paper selection process began with a preliminary review of the abstract and title of each paper to determine its potential relevance. This step allowed for an initial narrowing down of the pool of papers. Following this, we conducted a quick scan of the full paper, paying close attention to sections such as the introduction, methodology, and conclusion, to further assess its alignment with the inclusion criteria. Papers deemed relevant during this phase proceeded to a detailed review, where we thoroughly evaluated the study's objectives, methods, results, and contributions to ensure they met our requirements. This multi-stage process ensured that only the most relevant and high-quality papers were selected for inclusion in our study.

**Table 1** Inclusion and Exclusion Criteria.

| IC | Description Inclusion Criteria |
|---|---|
| IC1 | Studies analyzing ML pipelines within the Software Engineering (SE) domain. |
| IC2 | Papers proposing new ML techniques tailored for SE applications. |
| IC3 | Papers suggesting improvements to the ML pipeline in SE. |
| **EC** | **Description Exclusion Criteria** |
| EC1 | Publications that are inaccessible for reading or data collection, either due to being behind a paywall or not available through the search engine. |
| EC2 | Publications that are not in English. |
| EC3 | Publications that fail to satisfy the inclusion criteria. |
| EC4 | Publications that are already in the list (duplicated). |
| EC5 | Studies focusing on specific SE applications without pipeline insights. |
| EC6 | Papers applying ML for SE without proposing enhancements for the ML pipeline or novel methods. |
| EC7 | Off-topic papers (e.g., studies not involving ML models for SE). |

## 3.3 Paper Review

We review and categorize the final set of papers based on the stages of the ML pipeline (see Figure 2). For each paper, we document the experimental setup, proposed approaches, and findings to synthesize advancements in ML pipeline techniques for SE. This review provides a thorough understanding of research progress and identifies gaps in the ML pipeline for SE.

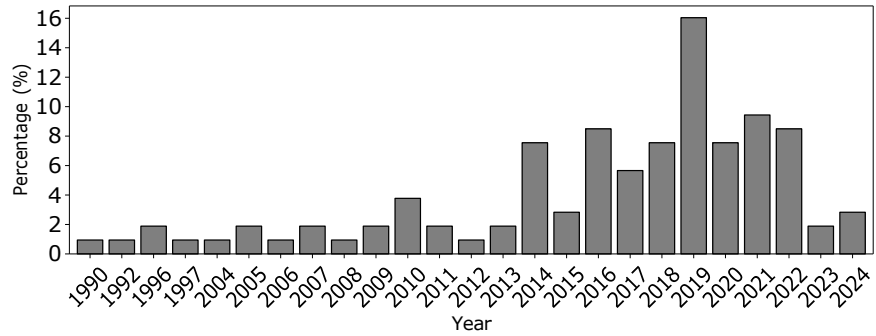

**Fig. 3** Outline of our methodology

## 3.4 Snowballing

To further ensure comprehensive coverage of relevant studies, we employed both backward and forward snowballing methods, as recommended in established guidelines [112, 55]. Forward snowballing involved reviewing the reference lists of the selected papers to identify additional studies that were cited by these works. This approach allowed us to uncover earlier foundational studies or related work that may not have been identified in the initial search. Conversely, backward snowballing involved tracing citations of the selected papers in more recent studies. This step ensured that we captured newer research building upon the selected works. By systematically applying these methods, we were able to expand our dataset with 20 additional papers specifically addressing ML pipeline improvements within the SE domain. This process not only enhanced the breadth of our study but also ensured the inclusion of both seminal and emerging research in this area.

## 4 Results

The yearly analysis of publication percentages reveals a steady growth in research output on ML pipelines for SE, particularly from the late 1990s onward. This trend reflects the increasing recognition of the importance of robust ML pipelines in addressing SE challenges such as defect prediction, code quality assessment, and effort estimation. The period from 2014 to 2024 is characterized by consistently high publication activity, with a notable peak in 2019, indicating heightened focus on optimizing and tailoring ML pipelines for SE-specific tasks. This sustained output suggests a thriving research community dedicated to refining ML pipeline stages, including data preprocessing, feature engineering, algorithm selection, and validation. The recent peaks demonstrate the field's responsiveness to evolving SE requirements and emerging technologies, underscoring the pivotal role of ML pipelines in driving innovation and improving software development processes in a rapidly changing landscape.

An analysis of publication types shows that journals dominate with 68.9% of the total publications, while conferences account for 31.1%. This preference for journals indicates a focus on detailed, rigorous, and comprehensive studies, which are characteristic of journal articles. Conferences, while vital for presenting preliminary findings and fostering rapid dissemination, appear secondary in this dataset. This trend reflects the emphasis on producing high-quality, peer-reviewed research over quick-turnaround conference papers, emphasizing the field's commitment to thorough scientific exploration.

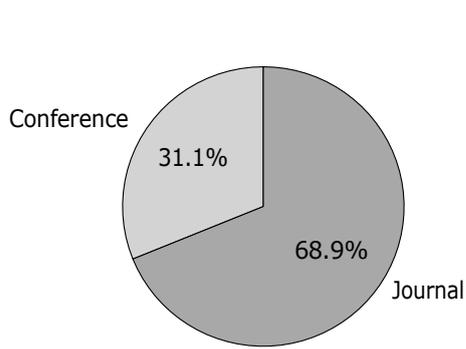 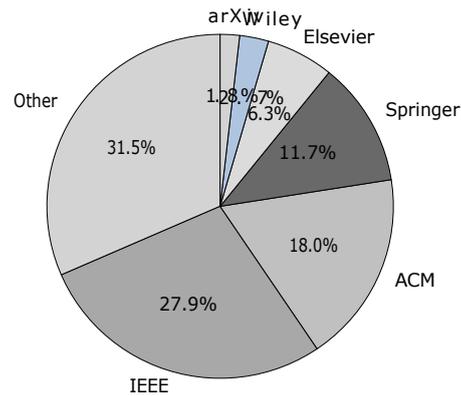

**Fig. 4** Percentage of Papers by Venue Type    **Fig. 5** Percentage of Papers by Venue Category

| Section | Citation | Perc (%) |
|---|---|---|
| Data Collection | [64], [11], [62], [92], [94], [49], [22], [76], [29] | 9% |
| Feature Selection | [37], [5], [77], [95], [82], [12], [4], [45], [52], [109], [3], [93], [113], [35], [75], [108] | 16% |
| Data Balancing | [97], [98], [17], [13], [63], [14], [1], [107], [105], [48], [115], [59], [61], [15], [32], [39] | 16% |
| Collinearity Analysis | [46], [44] | 2% |
| Noise Analysis | [40], [91], [114], [102], [84], [117], [101], [31] | 8% |
| Model Training | [66], [13], [30], [65], [85], [50], [57], [79], [41], [100], [80], [36], [90], [78], [43], [69], [25], [2], [24], [9], [83], [111], [10], [33], [26] | 26% |
| Model Evaluation | [96], [53], [70], [116], [110], [119], [58], [104], [42], [111], [23], [6], [34], [118], [41], [91] | 15% |
| Model Validation | [73], [99], [20], [21], [18], [16], [28] | 7% |

**Table 2** Distribution of papers across categories with percentages.

The distribution of publication venues reveals a strong preference for reputable platforms, with IEEE accounting for 27.9% and ACM for 18.0% of the total papers. These two venues alone represent nearly half of all publications, underscoring their central role in disseminating research within the field. Springer (11.7%) and Elsevier (6.3%) contribute smaller but significant proportions, while Wiley (2.7%) plays a more modest role. Notably, 31.5% of the papers fall into the "Other" category, indicating a broad spectrum of lesser-known venues. This distribution highlights the diversity in publication outlets while reaffirming the prominence of established publishers such as IEEE and ACM in maintaining high standards for academic contributions.

As shown in Table 2, the Model Training category constitutes the largest share (26%), reflecting the primary focus on developing robust and accurate ML models tailored to SE tasks. A significant number of studies in this field have trained multiple ML models to identify the most robust and effective approach for specific SE applications. Feature Selection and Data Balancing follow closely, each comprising 16% of the total papers. This distribution highlights the critical importance of optimizing feature sets and addressing class imbalance to improve model performance, both of which are foundational components of effective ML pipelines.

The Model Evaluation category, accounting for 15%, emphasizes a substantial focus on selecting appropriate evaluation metrics to ensure the reliability, generalizability, and practical applicability of ML models in SE contexts. In contrast, Model Validation (7%) and Collinearity Analysis (2%) receive comparatively less attention. These areas, while crucial, have been extensively studied and validated in broader ML contexts, reducing their prevalence as standalone topics within SE-specific research. The inclusion of Noise Analysis (8%) signals an increasing awareness of the detrimental effects of data quality issues on model performance. Studies addressing challenges such as class overlap, mislabeling, and sample duplication have demonstrated improvements in model accuracy and reliability, further underlining the importance of high-quality data in ML pipelines.

Overall, the table reveals that research in ML for SE is predominantly focused on core aspects of model development, including training, feature selection, and evaluation. While these foundational processes remain central, emerging areas such as noise mitigation and advanced validation strategies are gaining traction, indicating a gradual shift toward addressing more nuanced challenges in ML pipelines. This trend suggests that future research should aim for a more integrated approach, balancing efforts across underexplored areas to enhance the robustness, fairness, and interpretability of ML models in SE.

# 5 RQ1: What are the most effective data preprocessing techniques for improving the reliability of ML models in SE?

## 5.1 Data Collection

Researchers emphasize the importance of effective data collection for high-quality ML models [64]. Therefore, there are different sources and methods for data collection, what led to conduct several studies to select relevant data sources. For example, Bal and Kumar [11] proposed a method that uses the chi-square metric to identify the most relevant data sources. This approach showed that creating a heterogeneous dataset from different sources with many relevant metrics can improve the AUC[6] metric by at least 14%, with Random Forest being the most effective classifier. However, merging data from different sources can negatively impact performance if context and complexity are not taken into account [62].

---

[6]Area under the ROC Curve, the probability that the model differentiates between positive and negative cases

To study the impact of these data collection methods on defect prediction models, Sinaga et al. [92] compared 13 data collection methods from 61 projects using six ML classifiers. They found that each data collection method had a different impact on the project studied and on the model performance. In other study, to identify the source code changes responsible for introducing bugs into a software system that help to build the dataset, the SZZ method has been proposed with several variants such as Basic SZZ (B-SZZ), which considers all lines modified by a bug-fixing change [94], Annotation Graph SZZ (AG-SZZ), which discards non-semantic lines such as blank lines and comments [49], Meta-change Aware SZZ (MA-SZZ), which reduces the noise caused by branch or merge changes [22], and finally Refactoring Aware SZZ (RA-SZZ), which detects refactored lines and reduces modification-related noise [76]. Fan et al. [29] compared these variants to evaluate which one is better at capturing the data for defect prediction. They found that the variant used can lead to wasted effort, especially B-SZZ and AG-SZZ. In terms of interpretation, they found that the top-1 most important feature is always robust to the mislabelling noise. Contrarily to the second and the third most important metrics which are likely to be impacted except for Random Forest classifier.

**The effectiveness of merging heterogeneous data highlights the importance of metric diversity**, but only when carefully considering the context to avoid performance degradation. Methods like SZZ, along with its variants, allow for the refinement of data collection by isolating bug-inducing changes more precisely, reducing noise, and focusing on relevant modifications. **The variance in results among different classifiers and feature rankings further emphasizes the need to adapt data collection strategies to the chosen ML model**.

## 5.2 Feature Selection

Practitioners use feature selection methods to identify a relevant subset of features that accurately represent the underlying data distribution while reducing data dimensionality. Ghotra, McIntosh, and Hassan [37] found that the choice of feature selection technique wields substantial impact on the performance of models in the domain of defect prediction, an area extensively explored in ML for SE [5, 77, 95, 82, 12, 4]. However, Jiarpakdee, Tantithamthavorn, and Treude [45] reported that up to 94% of the selected metrics are inconsistent across feature selection methods. Also, Krishnan Rajbahadur et al. [52] found that classifier-specific and classifier-agnostic feature importance methods do not always agree with each other. Furthermore, feature interactions only affect feature importance for classifier-specific methods. This highlights the importance of carefully selecting the best method for selecting features. As a result, researchers have compared different methods of feature selection to determine which ones improve prediction quality. For instance, Wang, Khoshgoftaar, and Napolitano [109] and Ali et al. [3] conductive a comparative studies of different feature selection methods and found that the ANOVA method is the best method for best classification performance.

Due to differences in context and project complexity, using existing methods for feature selection may not always be efficient. Consequently, researchers have proposed innovative approaches to selecting relevant metrics for building ML models in SE. For example, Sinayobye et al. [93] proposed a hybrid classification model for feature selection based on correlation filter and ML classifiers on data that has a large amount of noise. They found that the random forest algorithm is the best for the proposed feature selection method. In other study, Xia, Yan, and Zhang [113] proposed an optimized support vector machine algorithm to analyze the importance of dataset metrics and found an improvement in accuracy and error rate performance. In other hand, Gao et al. [35] used a complex network of features to study the impact on the defect prediction model and found that complex network features give a better prediction model compared to traditional code features. Nam, Pan, and Kim [75] proposed a transfer learning approach to select relevant metrics for building defect prediction models. Their results demonstrated that their approach outperformed traditional feature selection methods in terms of accuracy.

Finally, Vescan, Găceanu, and Şerban [108] replicated prior research on cross-project defect prediction using composite classifiers and investigated the impact of different data preprocessing methods, including normalization, standardization, and feature selection. The study confirmed the original findings using raw data and found that performance improved with preprocessing techniques, particularly normalization combined with feature selection. Their results highlight that MaxVoting achieved the highest F-measure, while BaggingJ48 was the most cost-effective classifier.

**Overall, selecting an appropriate feature selection technique is an essential step in effectively selecting variables for software metrics analysis. The feature selection method can significantly impact the performance of the model.** Researchers have compared and recommended various techniques based on the specific context and application, and it is essential to carefully consider the trade-offs and limitations of each method. A thorough understanding of feature selection techniques can lead to improved decision-making in software metrics analysis, interpretation and ultimately enhance the quality of the ML model.

## 5.3 Data Balancing Techniques

Data balancing is essential in ML, especially for software defect prediction, where imbalanced datasets can bias models toward the majority class, reducing accuracy for the minority class. Various balancing techniques, including undersampling, oversampling, and synthetic sampling, are used to improve model performance on minority classes.

**As shown in Table 3, among the balancing methods, SMOTE and its variants (e.g., SMOTUNED, Borderline-SMOTE, and MAHAKIL) are the most frequently used**, indicating their effectiveness in addressing data imbalance in this domain. Techniques like Random Oversampling (ROS) and Random Undersampling (RUS) are also commonly used, particularly when computational efficiency is prioritized. Advanced methods, such as MAHAKIL and KMFOS, are introduced to enhance data diversity, showcasing innovation in balancing algorithms.

**Certain algorithms appear more frequently in combination with specific balancing techniques.** For instance, Random Forest (RF) and Support Vector Machine (SVM) are commonly used with SMOTE variants due to their robustness and performance stability across balanced datasets. Decision Trees (DT) and k-Nearest

Neighbors (KNN) also appear frequently, particularly when undersampling techniques, like Tomek Links or Condensed Nearest Neighbors, are applied, as these classifiers benefit from balanced datasets that reduce bias in prediction. Ensemble methods, such as EasyEnsemble and RUSBoost, show strong performance, particularly with imbalanced datasets where traditional undersampling fails, highlighting the advantages of combining balancing with ensemble learning.

The studies show that indicates that the selection of balancing technique can be tailored to both the classifier and the dataset's specific imbalance characteristics, offering targeted improvement in predictive accuracy. In summary, data balancing techniques are critical for ensuring fair class representation in ML models for SE. Addressing imbalanced data with these methods enhances accuracy, reliability, and interpretability of models across various SE tasks.

**Table 3**: Overview of Data Balancing Techniques and Their Impact on ML Models for SE

| Citation | Study Focus | Results |
| --- | --- | --- |
| Song, Guo, and Shepperd [97] | Analyzed 27 imbalanced datasets using 17 rebalancing methods across 7 classifiers. Evaluated with metrics like F1, MCC, G-mean, and AUC. | Medium and high imbalance levels negatively impact performance; models sensitive to data imbalance performed best. |
| Tantithamthavorn, Hassan, and Matsumoto [98] | Investigated various rebalancing techniques (SMOTE, ROSE, under-sampling) on 7 classifiers for defect prediction. | Significant performance improvement noted with under-sampling and logistic regression; AUC remained unaffected. |
| Çakır et al. [17] | Compared SMOTE, Random Oversampling (ROS), and Random Undersampling (RUS) on classifiers like Random Forest and Extra Trees. | ROS with Extra Tree Classifier yielded best accuracy (82%) and high F1-measure. |
| Bansal and Jain [13] | Explored undersampling methods (e.g., Tomek Link, Cluster-Based) using Decision Tree, K-Nearest Neighbor, and Naive Bayes. | Decision Tree performed best across methods; Cluster-Based undersampling was consistent. |
| Liu et al. [63] | Compared resampling, spread subsampling, and SMOTE with classifiers like C4.5 and SVM for defect prediction. | SVM and deep learning showed highest consistency across datasets; balance significantly influenced model outcomes. |
| Benala and Tantati [14] | Evaluated four oversampling methods (SMOTE, ADASY, SL-SM, SVM-SMOTE) with ensemble and decision tree classifiers. | SVM-SMOTE outperformed all others in accuracy enhancement across nine datasets. |

| Agrawal and Menzies [1] | Compared SMOTE vs. SMO-TUNED (automatically tuned SMOTE) using multiple classifiers, including Random Forest, Logistic Regression, and SVM. | SMOTUNED improved AUC and recall by 60% and 20% respectively across all classifiers. |
|---|---|---|
| Van Hulse, Khoshgoftaar, and Napolitano [107] | Compared repetitive under-sampling techniques: random undersampling, EasyEnsemble, partitioning, and RUSBoost. | RUS and baseline learners were less effective than the remaining techniques. |
| Tyagi and Mittal [105] | Assessed rebalancing approaches (e.g., SMOTE, MWMOTE, ADASYN) with SVM, KNN, and NN algorithms on multiple datasets. | Sampling techniques consistently improved model performance; ADASYN was the most effective. |
| Kamei et al. [48] | Compared ROS, SMOTE, RUS, ONESS for fault-prone module detection with LDA, LRA, and NN. | Sampling improved LDA and LRA performances significantly, with varied impact on other models. |
| Yu et al. [115] | Investigated cross-company data imbalance with filtering (NN, DBSCAN) and sampling (SMOTE, RUS) on 15 datasets. | NN filter prior to RUS yielded best results for all classifiers tested. |
| Li et al. [59] | Proposed data distribution approach for defect prediction; evaluated using KNN, C4.5, RF, and SVM. | Technique reduced imbalance impact, improving accuracy in defect prediction. |
| Li et al. [61] | Proposed DSSDPP, a domain programming-based approach for cross-project defect prediction, using NN-filter, TCA+, SSTCA+ISDA, BDA. | Achieved better results in MCC and AUC compared to other balancing methods. |
| Bennin et al. [15] | Introduced MAHAKIL, based on chromosomal inheritance theory, compared with SMOTE, Borderline-SMOTE, ADASYN, etc., on 20 defect datasets. | MAHAKIL improved recall and performance across classifiers, excelling with RF and KNN. |
| Fonseca and Bacao [32] | Proposed Geometric-SMOTENC, a sampling algorithm that handles nominal and continuous features, tested against four other methods on 20 datasets. | Geometric-SMOTENC outperformed other methods, improving G-mean and F-score significantly. |
| Gong, Jiang, and Jiang [39] | Compared balancing techniques on ML classifiers; proposed KMFOS, which clusters data and interpolates to generate missing class instances. | KMFOS yielded the best performance across classifiers, demonstrating superior balancing effects. |

## 5.4 Collinearity analysis

The literature studied the impact of different collinearity methods. **The researchers suggest practitioners to run correlation, redundancy analysis for a better interpretation of models and performances improvement.** Jiarpakdee et al. [46] studied the redundancy for the defect prediction models using 100 datasets. They found that 10-67% of metrics are redundant. Also, they found that the aggregated metrics can increase the redundancy. Additionally, Jiarpakdee, Tantithamthavorn, and Hassan [44] analyzed the impact of the correlated metrics on the interpretation of defect models. They used 14 defect dataset, logistic regression and random forest as learners. Annova one and two, scaled and non-scaled gini, scaled and non-scaled permutation for the interpretation. AUC, F-M and MCC as measures of performances. For each dataset, they removed the correlated metrics, interpret the model with and without correlation and calculate the impact. The obtained results showed that correlated metrics have the largest impact on the consistency, level of discrepancy, direction of the ranking of metrics, especially ANOVA. All studied datasets are improved by 15-64% for top-ranked metrics. Removing the correlated metrics improve consistency of ranking impact of the model performance less than 5%. In this paper, it was recommended to mitigate the correlated metrics especially for ANOVA, and avoid ANOVA one when correlated metrics are all removed.

## 5.5 Noise analysis

**The performance of ML models can be significantly reduced by noisy data**, leading to numerous studies that focus on analyzing and estimating the effects of various types of noise on these models [40] [91]. Yi et al. [114] studied the impact of class overlapping on the performances of different classifiers.Propose a cleaning method by combining improved K-Means clustering cleaning approach and Tomek-link method [102]. They used 6 open source projects. They compared the performance of LR, RF and KNN classifiers with and without cleaning. They used IKMCCA or KMCCA or NCL to estimate the impact on the used performances which are balance, recall, and AUC. the obtained results showed that this approach for removing the class overlap has a positive impact on all the performances for the used classifiers.

Rajbahadur et al. [84] studied the impact of the discretization noise of the dependant variable on ML classifiers. In this approach, they used 7 SE datasets which have a threshold confusion. They used RF, KNN, LR, and CART for classifications. They used Accuracy, Precision, Recall, Brier score, AUC, F-Measure and MCC. They varied the discretization noise zone of each case and estimate the impact on the performances and on the metrics importance. The obtained results showed that the impact of the discretization noise differs from a case to the other. In terms of metric importance, the results showed that the top three metrics are not impacted by the discretization noise. Rajbahadur et al. [84] provided a framework that allows the researchers to estimate the impact of the discretization noise on their own case of study.

Zhao et al. [117] studied the impact of sample duplication in ML based android malware.They used 4 datasets with duplicated samples. They used KNN, SVM, DT, RF for the prediction. To evaluate the model, they used precision, recall and f1 score.

To validate the models, they used 10 cross validation method. They found that sample duplication has an impact on the performance of unsupervised ML models, for the supervised ML models the impact is almost insignificant. In general, they found that the impact of sample duplication is undefined, it differs for each case of study.

Tantithamthavorn et al. [101] studied the impact and the nature of the mislabelling in ML. They injected randomly label noise to classification. Then, they manually created 3931 issues from Apache Jackrabbit and Lucerne systems and random forest for classification. They used Precision Recall, and F-Measure to measure the performances. They observed that issue report mislabelling is not random. In terms of performance, they found that the precision is rarely impacted, so it was recommended to rely on accuracy. Also, models trained on noisy area trained achieve only 56 - 68% of recall of models with clean data. The important metrics with the proper models are almost important on the non-clean models, except the second and third, which are instable.

Folleco et al. [31] studied the impact of the class noise on the random forest (RF100) classifier compared to C4.5, and Naive Bayes. They developed a new method for noise injection which has been used to inject different percentages of noise in the data. They measured the performances and calculate the impact. The obtained results showed that the class noise impacts the performances, so it is important to revise it. In addition, they found that the random forest model is the best model to apply with noisy class compared to the other models.

> *Effective preprocessing combines data collection, feature selection, data balancing, and noise handling to enhance ML models in SE. Diverse data sources and SZZ variants help refine data, while feature selection methods like ANOVA improve accuracy. Balancing techniques, such as SMOTE variants with Random Forest, reduce prediction bias on minority classes. Noise filtering and collinearity management further boost model stability and interpretability. Together, these tailored preprocessing steps lead to more reliable and accurate predictions.*

# 6 RQ2: Which ML algorithms demonstrate the best performance for SE tasks?

In the realm of ML (ML), experts are continuously creating and evaluating diverse ML models and approaches to enhance their effectiveness in SE tasks [66]. Various research studies have compared the efficiency of current models and introduced novel methods. Bansal and Jain [13] studied the performances of ML classifiers when under-sampling techniques are applied. They used Cluster Based, Tomek Link and Condensed Nearest Neighbours to fix under-sampling problem. For classification, they used K-Nearest Neighbour, Decision Tree and Naive Bayes. As measures, they used AUC, Precision, recall and F-measure. The results showed that Decision Tree is the best model with the applied techniques for under-sampling.

To identify the reasons for the poor quality of ML models, Fenton, Neil, and Square [30] studied the different software predictions models. The measures of token models are size, complexity metrics, quality data, testing metrics. By analyzing the obtained results, they found that the reasons of the bad quality are: the missing of data variety and specification, the use of the inappropriate data to build the dataset, the modelling mistakes. They also observed that optimizing the complexity and the size of the learning model are not sufficient to have good performances. consequently, when the model is bad, the obtained results are not easily explainable.

Lomio, Moreschini, and Lenarduzzi [65] studied the combination between statistical code analyser and ML model building. They used 29 Java project with 58125 commit with 33865 faults and the SZZ algorithm to collect the data. for data balancing, they used SMOTE technique. To collect the metrics, they used SonorQube tool, they collected 174 variables. As ML models, they used RF, gradient Boosting and XG Boost. As deep learning models deep learning, they used residual network and fully conventual network. The obtained result showed that deep learning models perform better than ML model in terms of AUC. Comparing to the basic models, combining Sonarqube metrics with good predictors do not show a significant improvement, especially for deep learning.

Rodriguez et al. [85] proposed to use descriptive approach for subgroup discovery instead of classification. They used two subgroup discovery algorithms, SD [50] and CN2-SD [57], the difference between the algorithm is that the first one takes many rules, and the second one can resist to the overlap. They used data merging to combine multiple analysed datasets in 3 new ones. The obtained models show good performances in terms of precision. The results showed that these algorithms can be robust to detest the most defectives modules in the system and also in the case of highly imbalanced data, noisy data, redundancies, and inconsistencies.

Pérez-Verdejo, Sánchez-García, and Ocharán-Hernández [79] presented a systematic literature review about ML for automated requirement classification. They found Naive Bayes, J48, and K Nearest Neighbour classifiers are the most used models for the classification of software requirements. For the datasets, they found that almost used from academic databases. For the evaluation metrics, precision recall and F1 measure are generally used to evaluate the performances. Additionally, Hernández-Molinos, Sánchez-García, and Barrientos-Martínez [41] conducted a systematic literature review that study the performances of different classifiers in defect prediction. They found that Naive Bayes and Random Forest are the most used models for defect prediction.

Tantithamthavorn et al. [100] studied the optimization of the parameters of 10 classifiers for defect prediction. They modified different parameters for each algorithm. For example, the number of the neighbour of KNN algorithm. For validation, they used the bootstrapping technique. They found that caret is the most stable algorithm, where the performances by 40 percent in terms of AUC. In addition, caret increases the likelihood of producing a top-performing classifier by 83 percent. Generally, they concluded that the parameters of classifiers impact the model performances, so they needed to be investigated.

Petrič et al. [80] proposed an approach for combining classifiers from different families to build one prediction model on the same dataset, which called build stacking ensemble. They used Naïve Bayes, C4.5 decision tree, K-nearest neighbour, and sequential minimal optimization. For each model, they regularized the parameters. They used 8 datasets. The obtained results showed stacking ensemble technique improve the performances in terms of accuracy. Also, they noticed that the diversity between the used classifiers is important to improve the performances. Finally, the stacking ensemble could reduce the number of misclassifications.

Ge, Liu, and Liu [36] studied the supervised algorithms with imbalanced data. They used for 8 datasets of NASA. For classification, LWL, C4.5, Random forest, Bagging, Bayesian Belief Network, Multilayer Feed forward Neural Network, SVM and NB-K algorithms were used. To balance the data, they used cost-sensitive learning is used to reset instance weights. Consequently, each class will have the same total weight. As metrics, they used FP rate, MCC, F-measure, Precision, and AUC. The obtained results showed that bagging combined with Random forest classifier has good performances in terms of recall, and large classification capability compared to the others algorithms.

Schroeder et al. [90] studied the impact of the model size on the rapidity of prediction and software. They compared the performances of artificial neural networks (ANN), support vector regression (SVR), long short-term memory (LST), autoregressive integrated moving averages (ARIMA), and Holt's linear trend method (HOLT) in long and short term size in an industrial context. The obtained results showed that the impact differ significantly from case to another. Also, they noticed that the SVR model was the worst model in this case of study.

Panichella, Oliveto, and De Lucia [78] analyzed the equivalence of different defect predictors and proposed a combined approach CODEP which takes regression and Bayesian based models. As models, they used logistic regression, bayesean network, Radial Basis Function Network, Multi-layer Perceptron, Alternating Decision Trees, Decision Table. To measure the equivalence between the models, they used PCA, which identifies orthogonal dimensions of the models. They also used 10open source software systems to build the dataset, precision and recall, AUC to measure the performances. The obtained results showed that the investigated classifiers are not equivalent, and they can complement each other. The overlap between the models is low. They also found that CODEP perform the best then the other alone models on the used performances.

Herbold, Trautsch, and Grabowski [43] replicated a study to compare local and global model for cross project prediction where the data is coming from different sources. Local models techniques try to find homogeneous regions of the data. Then, train each classifier on the suitable region. To build local models, they used WHERE [69], N-WHERE which uses WHERE with normalized data, EM [25], N-EM which uses EM with normalize data. Ad global model, they used KNN. As metrics, they used recall, precision, F-measure and error rate.The obtained results showed that local model make a minor difference compared to global models and transfer learning for cross-project defect prediction.

Ahsan and Wotawa [2] studied the impact of the multi-labelling when having resolved software change requests (SCRs). In this case, it is possible to predict the coming file changes. They used Latent Semantic Indexing (LSI)[24] to index the key terms of SCRs. For data, they used datasets from the three OSS projects repositories which are Mozilla, Eclipse, and Gnome. To measure the performances, they used recall and precision. They obtained 58.2% and 47.1% precision for single and multi label. For individual label, they obtained 86.5% and 92% in terms of precision.

Aydin and Tarhan [9] developed a model to predict defect arrival phases when historical data is unavailable. They compared a Linear Regression model with the Rayleigh model [83], using data from 29 consecutive project releases and 15 module-level releases. Performance was evaluated using root-mean-square error (RMSE), mean absolute error (MAE), and mean magnitude relative error (MMRE). Results indicated that the Linear Regression model outperformed the Rayleigh model on both RMSE and MAE. Additionally, the Linear Regression model achieved an MMRE below 0.25, while the Rayleigh model also demonstrated acceptable performance with an MMRE below this threshold.

Wang et al. [111] showed in their SLR that NB, SVM, RF, DT, RF, LR and KNN are the most used models for SE tasks. Among those tasks, they mentioned Requirements Detection and Classification, Code Smell/Anti-pattern Detection, Test Automation and Prioritization, Defect Prediction, Defect Localization, Sentiment Analysis, Code Clone and Similarity Detection, Software Quality Prediction and bug related prediction.

Bahel, Pillai, and Malhotra [10] compared different ML classifiers with and without hyperparameters, which are logistic regression, naive Bayes, K-nearest neighbours, decision tree and random forest. They used UCI ML Breast cancer Wisconsin's dataset [33] and titanic dataset from disaster [26] and accuracy as metric. They found that the random forest classifier is the most optimal classifier followed by naive Bayes compared to the other classifiers.

**The studies highlight that different ML algorithms exhibit varying strengths across SE tasks, often influenced by specific characteristics of the data and preprocessing techniques applied.** For example, ensemble methods like Random Forest and Decision Trees consistently deliver strong results, underscoring their robustness in handling class imbalances. Additionally, stacking ensemble approaches that combine diverse classifiers, such as Naïve Bayes, K-Nearest Neighbors, and Decision Trees, have been shown to improve prediction accuracy by leveraging the unique strengths of each model. Deep learning models, such as residual networks, perform well in scenarios with larger datasets but offer minimal performance gains when integrated with additional code metrics. On the other hand, simpler models like Naïve Bayes and Logistic Regression are still widely used for their interpretability and efficiency, particularly in requirements classification and other straightforward SE tasks.

**Moreover, studies on hyperparameter tuning reveal that optimized parameters significantly enhance classifier performance, suggesting the necessity of tuning for high-stakes tasks like defect prediction.** Specialized approaches, like subgroup discovery algorithms and local models, show promise for

handling noisy, imbalanced, or complex datasets, while global models remain effective for cross-project prediction.

> *Random Forest and Decision Trees emerge as versatile and high-performing algorithms for defect prediction tasks, with ensemble methods offering substantial improvements in accuracy and resilience across various SE scenarios.*

# 7 RQ3: What evaluation metrics most accurately assess the effectiveness of ML models in SE applications, and how do these metrics impact model selection?

Selecting appropriate evaluation metrics is crucial for accurately assessing ML models in SE contexts. Numerous studies have explored diverse metrics to enhance model evaluation quality and accuracy [96, 53, 70, 116, 110, 119, 58, 104, 42, 111]. For example, Song and Minku [96] examined the effect of evaluation waiting time on just-in-time (JIT) defect prediction models, using F-measure, precision, and G-mean. They observed that shorter evaluation waiting times positively impact evaluation results, suggesting that timeliness in model evaluation is a factor worth considering in defect prediction.

Similarly, Mori and Uchihira [70] introduced a novel classification model, superposed Naive Bayes, evaluating its performance against other classifiers based on accuracy, AUC, and interpretability. Their findings indicated that their model achieved a balance between AUC (for accuracy) and model interpretability, highlighting that AUC can be an effective measure for models that need to balance predictive power and comprehensibility.

Metrics also play a pivotal role in feature selection, as Wang, Khoshgoftaar, and Napolitano [110] showed. Their study found that the Best Arithmetic Mean (BAM) metric within wrap-based feature selection produced optimal results across datasets with varying imbalance levels. This suggests that BAM may offer a robust approach to feature selection, particularly when handling imbalanced data.

Tunkel and Herbold [104] explored the connection between various performance metrics and the cost implications of ML models, using 20 performance metrics along with Herbold [42]'s model to evaluate cost. They found that metric groups often correlate and that certain classifiers, like LR, DT, and RF, carry a considerable risk of suboptimal cost efficiency, underscoring the importance of metric selection when optimizing for cost-related goals.

Explainability remains a key concern in model evaluation. According to Dam, Tran, and Ghose [23] and Allahyari and Lavesson [6], decision trees are often seen as the most understandable ML models in SE due to their interpretable nature. However, as Freitas [34] caution, decision trees can overfit training data and may fail to capture complex relationships, particularly as tree size increases. Additionally, Zheng, Shen, and Chen [118] reported that enhancing interpretability, as seen in their JUST model for JIT defect prediction, can also improve model accuracy. JUST, which combines traditional statistical and deep learning approaches, demonstrated superior accuracy and interpretability.

A comprehensive literature review by Wang et al. [111] identified accuracy, precision, recall, F1, AUC, Mean Average Precision (MAP), and Mean Absolute Error (MAE) as commonly used metrics across classification and regression tasks. Hemández-Molinos, Sánchez-García, and Barrientos-Martínez [41] further emphasized that precision and recall are most frequently applied in model evaluation, especially for defect prediction. Additionally, Sharma et al. [91] found that AUC remains a preferred metric in defect prediction, particularly valued for its ability to summarize model performance across thresholds.

The reviewed studies underscore that metric selection is not one-size-fits-all; instead, it should align with the specific goals of the model, whether that is accuracy, interpretability, or cost-efficiency. AUC, precision, recall, and F1 are commonly used due to their balance between sensitivity and specificity, making them ideal for tasks like defect prediction where misclassification costs are high. Furthermore, metrics like BAM and cost-based evaluation metrics have emerged as valuable for feature selection and cost-sensitive applications, respectively, indicating that customized metrics may optimize model performance under certain conditions. Explainability, often tied to decision trees and interpretable models, proves essential in SE contexts, particularly when complex relationships must be communicated to stakeholders.

> *Choosing the right evaluation metrics, such as AUC, precision, recall, and F1, is essential for accurately assessing ML models in SE, especially in tasks like defect prediction where balancing sensitivity and specificity is crucial. Additionally, metrics tailored to specific goals—like interpretability for decision trees or cost-efficiency—enhance model selection, making evaluation metrics as important as the models themselves for effective application.*

## 8 RQ4: How do different model validation techniques affect the stability and generalizability of ML models in software defect prediction?

In ML, the choice of validation technique plays a crucial role in determining model suitability, stability, and generalizability [73]. Tantithamthavorn et al. [99] compared various ML model validation methods using 101 public defect datasets and identified a 77% chance of unstable results across these datasets. They evaluated 12 commonly used validation techniques, including variants of holdout, bootstrapping, and cross-validation, and used a range of metrics: threshold-dependent and threshold-independent performance measures, precision, recall, AUC, Brier score, and calibration slope. Ranking the validation methods with the Scott-Knott Effect Size Difference Test, they found that ordinary bootstrapping is the most stable validation technique across metrics and classifiers, while single repetition holdout exhibited high bias and variance, making it less reliable.

Cito et al. [20] introduced a model-agnostic technique to interpret mispredictions in ML models, using labeled datasets in multiple scenarios to guide actions such as improving training data or adjusting model choice. This rule-learning algorithm, which takes a dataset, model, and target as inputs, provides explanations for mispredictions. Compared to methods like RIPPER [21] and STUCCO [18], their approach demonstrated superior interpretability and effectiveness in explaining model errors.

The fairness of model validation pipelines has also been investigated. Biswas and Rajan [16] studied data transformers in different ML pipeline stages across 37 pipelines, using fairness metrics like Statistical Parity Difference (SPD), Equal Opportunity Difference (EOD), Average Odds Difference (AOD), and Error Rate Difference. Their results showed that certain preprocessing steps, such as data filtering and handling missing values, can skew data distribution and introduce bias. Additionally, techniques like feature standardization and non-linear transformations were found to improve fairness, while feature selection and sampling methods risked reducing fairness, indicating that the choice of validation pipeline components directly impacts model fairness.

Moreover, Falessi et al. [28] analyzed the effect of data order on validation, specifically comparing validation techniques that preserve data sequence, such as walk-forward validation, 10-fold cross-validation, and bootstrapping. Using nine classifiers on 15 datasets from industry and open-source projects, they found that the walk-forward technique statistically outperformed cross-validation and bootstrap in terms of AUC, precision-recall, and MCC. Notably, they observed an 83% difference in defect prediction accuracy between the first and second halves of the dataset, underscoring the importance of temporal order in model evaluation.

**The reviewed studies underscore that model validation methods significantly impact both the stability and generalizability of ML models in defect prediction.** Bootstrapping stands out as a stable choice for achieving consistent results across a wide range of performance metrics, while walk-forward validation excels in contexts where temporal order is essential, such as real-world software release cycles. Techniques like data filtering, feature transformation, and sampling, however, can introduce bias and fairness issues if not carefully tailored to the model and dataset, emphasizing the need for context-aware validation approaches. Together, these insights suggest that selecting appropriate validation techniques requires careful consideration of dataset characteristics, model requirements, and fairness concerns to ensure robust and reliable defect prediction outcomes.

> *Our findings indicate that validation techniques like bootstrapping provide stable, generalizable results across metrics, while methods like walk-forward validation are particularly effective when temporal order is crucial. Additionally, choices in preprocessing and validation steps, such as feature transformation and sampling, must be carefully managed to avoid introducing bias, underscoring the importance of context-aware validation for reliable defect prediction.*

# 9 Conclusion

This systematic literature review investigates the role of machine learning (ML) pipelines in addressing critical challenges within software engineering (SE), including defect prediction, code review, and quality estimation. The findings highlight the importance of robust preprocessing techniques such as SMOTE for data balancing and advanced feature engineering methods like SZZ-based algorithms for bug identification. Ensemble methods, such as Random Forest and Gradient Boosting, consistently deliver superior performance across SE tasks, while deep learning models excel in large datasets with complex patterns. Simpler models, such as Naïve Bayes, maintain relevance for their interpretability and efficiency in specific scenarios. Evaluation metrics like AUC and F1-score dominate performance assessments, but emerging metrics such as Best Arithmetic Mean (BAM) show potential in specialized applications. Bootstrapping is identified as a stable validation technique, whereas walk-forward validation is crucial in temporal data contexts like release cycles. Despite significant progress, challenges persist, particularly in achieving end-to-end integration of pipelines, managing imbalanced and noisy datasets, and adapting models to domain-specific contexts. This review provides actionable insights for researchers and practitioners to refine their approaches, emphasizing the need for dynamic and context-aware pipeline designs that align with the evolving demands of SE. Future work should focus on bridging the identified gaps, fostering the development of unified frameworks that optimize ML applications for increasingly complex software systems.